# CHINESE HERB MEDICINE IN AUGMENTED REALITY

*Qianyun Zhu[1] · Yifeng Xie[1,2] ·Fangyang Ye[1,3] ·Zhenyuan Gao[1,5] · Binjie Che[5] · Zhenglin Chen[1]· Dongmei Yu[1]*

1 School of Mechanical, Electrical & Information Engineering, Shandong University, Weihai, Shandong, China
2 Center of Precision Medicine and Healthcare, Tsinghua-Berkeley Shenzhen Institute, Shenzhen, Guangdong, China
3       Department of Automation, Hangzhou Dianzi University, Hangzhou, Zhejiang, China
4       Department of Software Engineering, Northeastern University, Liaoning, Shenyang, China
5  Hanhai Xingyun Digital Technology Co., Ltd, Tianjian, China

**ABSTRACT**

Augmented reality becomes popular in education gradually, which provides a contextual and adaptive learning experience. Here, we develop a Chinese herb medicine AR platform based the 3dsMax and the Unity that allows users to visualize and interact with the herb model and learn the related information. The users use their mobile camera to scan the 2D herb picture to trigger the presentation of 3D AR model and corresponding text information on the screen in real-time. The system shows good performance and has high accuracy for the identification of herbal medicine after interference test and occlusion test. Users can interact with the herb AR model by rotating, scaling, and viewing transformation, which effectively enhances learners' interest in Chinese herb medicine.

**INTRODUCTION**

Plants are always the key source of drug or treatment strategy in different traditional medicine systems. With the advent of the industrial revolution and the introduction of modern drugs, the use of herb was abandoned for a specific period of time[1]. However, the obstacles on the way of natural compounds studies have recently diminished mostly through using modern techniques. This has resulted in higher interest in using natural compounds in pharmaceutics[2, 3], and many people tend to choose plant-based medicines or products to improve their health conditions or as curative substances either alone or in combination with others. Herb medicine includes herbs, herbal materials (like plant parts) or preparations, processed and finished herbal products, active ingredient[4].

In recent years, the use of herbal product has beginning to revive hugely due to the of modern drugs, failure of modern therapies against chronic diseases, and microbial resistance[5-7]. Herb medicines are being used by 75-80% of the world population, especially those living in developing countries[8, 9]. In India, approximately 70% of the modern drugs are discovered from natural resources and the number of other synthetic analogs have been prepared from prototype compounds isolated from plants[10, 11]. The Thai government promotes and advocates the use of traditional and alternative health care modalities through scientific research and product development[12-14]. Chinese herb medicine (CHM) is accepted by people widely and develop with modern technologies. It was estimated that 11% of the total 252 drugs found in the essential medicine list of WHO are exclusive of plant origin[15, 16]. Multiple surveys reported that people with cancer commonly use herbs or herbal products to slow down the metastatic transition, supporting the immunity system and reducing stress[17].

The number of medicals with herbal medicines is increasing[18]. Global pharmaceutical companies and researchers equipped with modern scientific knowledge, technology, idea and started to rediscover medicinal plants as a source of new drug candidates based on traditional knowledge[19, 20]. People study the effects of herbs on animals and the relationship between herbs and prescription drugs so that they can be used in clinical treatments[21-23]. However, the learning and popularization of herbs are limited by their great variety and high similarity, which makes it difficult for beginners to remember and to develop an interest in learning them. As never before, the characteristics of Augmented reality (AR) suggest that it can solve this kind of learning difficulty[24]. In addition to the 2D and 3D objects, digital assets such as audio and video files, textual information, and even olfactory or tactile information can be incorporated into users' perceptions of the real world. Collectively, these augmentations can serve to aid and enhance individuals' knowledge and understanding of what is going on around them[24-26].

AR deriving from Virtual Reality (VR) enriches and renders the real world with digital information and media such as 3D models, which overlays in the camera view of a smartphone or connected glasses in real-time. AR is known to be a virtual object that is generated by a computer through the real environment seen by a mobile phone, tablet, or AR glasses[27]. It integrates digital or virtual information such as images, audio, video, and haptic sensations with the natural environment seamlessly. AR is regarded as the blend or the 'middle ground' between synthetic and real world[28].

The implementation of AR is defined by three characteristics: (a) the combination of real-world and virtual elements, (b) which are interactive in real-time, and which (c) are registered in 3D (i.e., the display of virtual objects or information is intrinsically tied to real-world loci and orientation)[29-31].

Human-computer interaction proceeds through smartphones or AR browsers as an emerging novel technology[32, 33]. AR has a broad spectrum of applications such as education, military, and medicine, which displays the contents informatively in any media such as video clips, animation, 3D models, etc[9, 34, 35]. AR technology mainly works through the identification of the target object, and then tracking the identified objects, after that imposed virtual images onto the tacked object which then present it by the display device[36, 37].

We apply AR to CHM with the goal of improving the learning and recognition of diverse herbs. There are thousands of Chinese herb plants, and we choose 88 types of plants as the initial learning objective, which are commonly used in medical treating and daily body caring. Those 88 plants cure a wide range of diseases including influenza, the coldness of body, renal dysfunction, and so on. They are highly representative in Chinese medicine because they have a great effect on body caring, which is the main topic in Chinese medicine. Herb in AR stimulates cognitive learning through the mobile application. The android application includes herb species information, morphological characteristics, ecological habits, and AR models. We can include more herb models in the future as a comprehensive learning platform for CHM.

## RESULTS

**Herbal Model Creation**. We collect photographs from all angles of the herbs in order to get a complete knowledge of their structure. Using 3ds Max to build 3D models of the actual shape of the herbs, and then we use V-Ray to render the models so that they are closer to the real material. The supplementary Figure.1 is a rendering and classification of all the herbs (Supplementary Figure.1).

**System and Platform.** CHM recognize system has the function of accurately identifying herbs and displaying the 3D structure and information of herbs. We use the mobile tablet terminal of the android system to test the display effect and usability of the herb learning platform. Tablet is more convenient than desktop, which has larger calculation capability than the mobile phone. The tablet is portable for location-independent learning and students could visualize 3D herb models to explore herb structural detail any time.

**Function test.** The application interface is user-friendly including the initial view, the recognition interface, and herb information (Figure.1). In the test phase, we use an external camera (TianTianQuan, JW-02, 1920*1080) to capture images. In fact, when users use a mobile device officially, it's the device's built-in camera that gets the image. When the camera scans the picture of CHM, the corresponding 3D AR herb model and its information will pop up on the screen.

The herb learning application has good interactivity that exposes knowledge about CHM to participants. Users can learn herb details by selecting the function buttons including herb species, morphological characteristics, and ecological habits (Figure.1c). The species information contains the herb's Chinese and English name, source area, and usage. Morphological characteristics describe the anatomy features including the roots, stems, leaves, and seeds. Ecological habits introduce the suitable growth environment and life cycle of the herb. All herb background knowledge coexist with the 3D herb model.

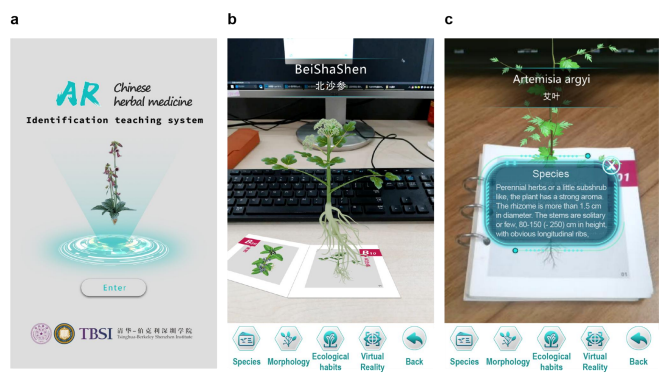

**Figure.1 The user interfaces of the application. a** The initial design. **b** Recognition interface. **c** Herbal information.

**Confusion test.** To test the effectiveness of the herb AR display module in the presence of visual interference, we conduct three controlled trials:

(i) Camera rotator

We rotate the camera angles by 45, 90, and 180 degrees respectively to verify the system performance after camera rotation (Figure.2a). The results show that the system is robust to camera rotation and the same 3D herb model appears without perturbation.

(ii) Occlusion test

We cover the picture with blank paper to show whether the system could identify the herb after losing 50% of the information (Figure.2b). The results are encouraging, where the partial picture is able to trigger the display of the whole 3D model. It's worth noting that when select one herb with a small area and less characteristic the system may not identify it (supplementary Figure.2b).

(iii) Interference test

We test the accuracy of the recognition module by placing other plants next to the correct herb picture. As shown in Figure.2c, the application can identify the correct herbal information.

(iv) Color test

We adjust the herb image to grayscale to verify that the system can still work after the image loses color information. As shown in Figure.2d, the test result is exactly as we expected.

All the experiments show that the module can overcome different visual interference. The system is robust to variation in the 2D image that increases the generalization and applicability of the platform.

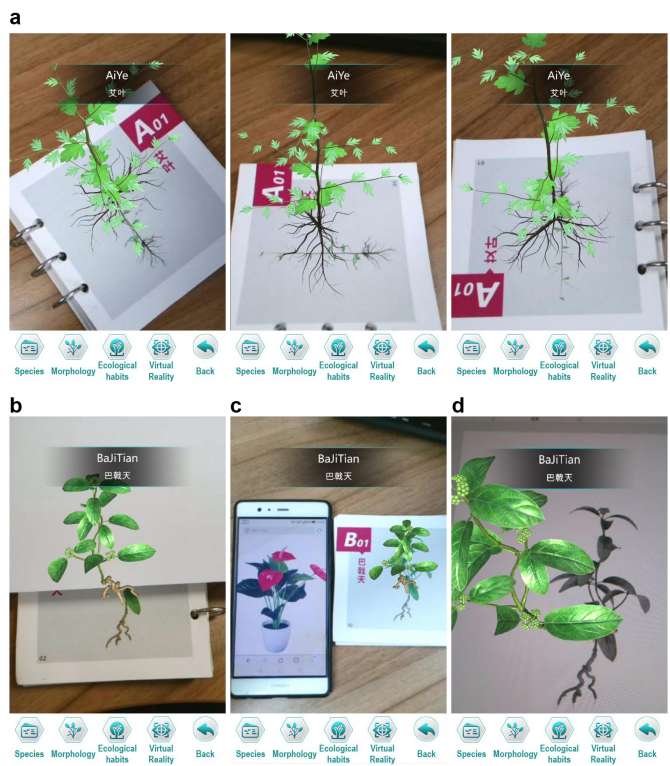

**Figure.2 Three kinds of visual interference tests. a** Rotate the angle by 45,90, and 180 degrees. **b** Occlusion test. **c** Interference test. **d** Color test.

## DISCUSSION

In recent years, AR technology becomes mature and make great achievements in various applications. With the maturity of educational information technology, AR technology has been gradually integrated into the field of education[38]. We develop an AR CHM teaching system, which helps the user to enhance his vision and interest in the herbs.

In our current work, we have only added 88 herbs commonly found in ten regions of China. In fact, there are about 12,000 kinds of medicinal plants in China, among which more than 5,000 kinds of Chinese herbs have been used[39]. For future works, we will add more herb medicine to our database and design more human-computer interaction features. Meanwhile, we will increase the information about the effectiveness of herbs and simplify all information to make it more readable and easier to remember. In addition, we also hope to increase in the database rather than herbal medicines, including animal drugs and mineral drugs. After system optimization, we will apply this herb-learning platform to real class to create a new learning environment that is interactive, interesting, imaginative, and intelligent.

## METHOD

AR is a collection of real-world and digital information. Users can visualize the virtual models superimposed with the real world and interact with the virtual environment.

**Model creation and system interface design.** We build the AR models with 3dsMax 2014 and Unity 2017 in the Android environment configured for the application export. The AR operations implement with the object-oriented programming C# package, The AR system consists of an autofocus camera and mid-range computer with a 3.6 GHz quad-core Intel i9-9900K processor and NVIDIA GeForce RTX 2080 graphics card.

We construct and render 88 3D models for commonly CHM contains Lingzhi, Aiye, Bajitian, and other 85 herb plants. Model building requires the tools of Edit Mesh and UVW Mapping in 3ds Max. The final representative model is demonstrated in Figure.3. The user interface (UI) scenarios and functionality of the AR application are realized in Unity. We create the cameras, canvas, lights, and objects as the scenes and associate the target object with scripts including text, picture, and button for user interaction through the Unity Engine class library from C#.

**The development of the application.** The display module for AR herb is the Vuforia SDK[40]. The Vuforia QCAR algorithm calculates the similarity between scanned and the target image from the database to trigger the display of the corresponding 3D models. The realization of the herb AR display module involves two stages (Figure.3).

Specifically, in the first stage (Figure.4a), we upload herb pictures that modeled by ourselves to the Vuforia target database and generate a unity package file. Second, we pack the total herb model by binding the file with the corresponding 3D virtual herb model in Unity. Besides, we unify all the herb information into a ".txt." file, which includes herb species, morphological characteristics, and ecological habits. Then we add the ".txt" file in the model as supplementary information. Third, we launch the herb model app on the android platform.

In the next stage (Figure.4b), when we show the herb image to the app, the app calculates the similarity between scanned image and target image from the Vuforia target database to trigger the corresponding 3D model display. And users can feel the herb visual effect in the real work after the bound 3D model rendered by Unity. Users can learn herb details such as herb species, morphological characteristics, and ecological habits in the app, where they interact with the 3D model by rotating and scaling freely.

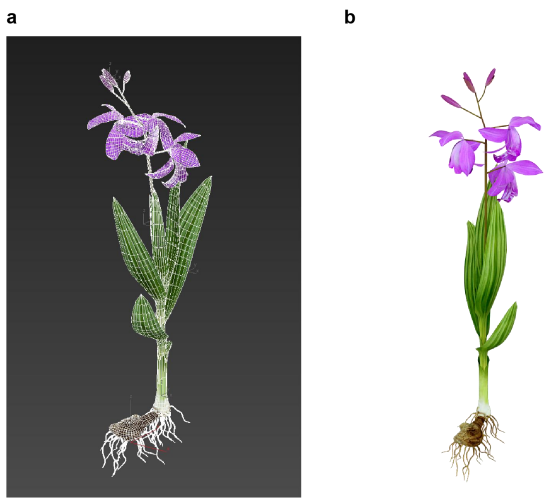

**Figure.3 A 3D model and rendering of the representative herb. a** 3D model of Bai Ji. **b** Rendered image.

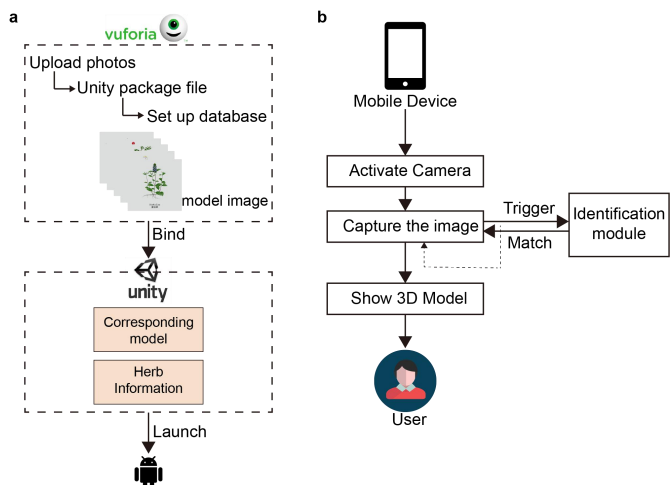

**Figure.4 The working principle of the AR platform. a** Identification module. The identification of herbs is realized through the Vuforia SDK, and then bind the corresponding model and information to the Unity. **b** The mobile camera captures the herb image and triggers the module, so the users can achieve the matching herb model and detail from the app.

**ACKNOWLEDGEMENTS**

We would like to thank Hanhai Xingyun Digital Technology Co., Ltd and Shiyao Zhai, Bin Pan, Dongheng Zhai, and Yanjiang Jia for the support of model construction, algorithm development, testing and feedback. This work is supported by National Natural Science Foundation of China (31970752), Science, Technology and Innovation Commission of Shenzhen Municipality (JCYJ20190809180003689, JSGG20200225150707332 , JSGG20191129110812), and Shenzhen Laboratory Open Funding (SZBL2020090501004).